\documentstyle[12pt]{article}
\begin{document}
\begin{titlepage}
  \hfill   OHSTPY-HEP-T-93007 \\
\vskip .1in
\center{\bf Deducing ${\cal L}_{fermion}$ at M$_{GUT}$ using Low Energy
Data}
\center{\bf or}
\center{\bf Towards a Theory of Fermion Masses}
\vspace{.5in}
\center{Stuart Raby~\footnote{Talk presented at
the HARC Workshop on Recent Advances in the Superworld,
Houston, Texas, April 14-16, 1993.}~\footnote{In
collaboration with Greg Anderson, Department of Physics, The Ohio State
University, Columbus, OH 43210; Savas Dimopoulos, Department of
Physics, Stanford University, Stanford, CA 94305; Lawrence J. Hall,
Department of Physics, University of California and Theoretical
Physics Group, Lawrence Berkeley Laboratory, 1  Cyclotron Road,
Berkeley, CA 94720 and Glenn Starkman, Department of
Physics, University of Toronto, Toronto, Canada}\\ {\it Department of
Physics, The Ohio State University\\ Columbus, OH 43210}}

\vspace{.5in}

\flushleft{\large\bf 1 Introduction}

In the last 20 years we have accumulated an enormous amount of data
on elementary particles and their interactions. This data serves two
purposes: to fix the phenomenological parameters of the Standard Model
[SM] and to verify that the SM is an excellent description
of nature.  It is our goal to understand the origin of these many
arbitrary parameters. In this talk we consider a supersymmetric [SUSY]
SO(10) grand unified theory [GUT].  We present a straightforward
procedure, incorporating a general operator analysis,  which allows
us to use low energy data to determine the fermionic sector of the
theory at the GUT scale~$^1$.   In what follows  we first
review the status quo, i.e.  the low energy data [LED].  We then
review the evidence for SUSY GUTs, and focus on the virtues of the
particular group SO(10).  Following a discussion of our dynamical
principles, we present a general operator analysis for ${\cal
L}_{fermion}$ ,i.e. the fermionic sector of the GUT theory.  We argue
that all fermion masses and mixing angles can be described with a
minimum of 5 arbitrary parameters in the Yukawa sector of the theory
at the GUT scale, $M_G$.  Including the parameter $\tan \beta$, the
ratio of Higgs vevs present in any SUSY theory, we thus obtain a 6
parameter description of fermion masses and mixing angles, leading to
8 predictions.  Finally, we present preliminary results.  These
preliminary results are encouraging and eminently testable.
Notwithstanding, the power of our analysis is in the paradigm whereby
the use of LED, hand in hand with symmetry conditions,  allows us
determine the fermion sector of the theory at $M_G$.
\end{titlepage}
\setcounter{section}{1}

\section{Status Quo}

There are 18 arbitrary parameters in the SM with 13 of these in the
fermion sector of the theory.\footnote{That is if we assume that
neutrinos are massless.  Incorporating neutrino masses leads to an
additional 9 parameters,  3 masses and 6 mixing angles.} Let us
briefly review the state of our knowledge of these 18 parameters
since they play a central role in what follows.  In Table 1  we list
the 18 parameters of the SM along with their experimental and/or
theoretical uncertainties.

\begin{table}[h]
\begin{center}
\begin{tabular}{|c|c|c|}
\multicolumn{3}{l}{Table~1. The 18 parameters of the Standard
Model.}\\
\multicolumn{3}{c}{}\\ \hline\hline
Parameter & Uncertainty & Comments  \\ \hline
$\alpha, \sin^2\theta_W$ &  $<$ 1/2 \%  &  high accuracy \\
$\alpha_s(M_Z)$ &  $\sim$ 10\%  &  less certain -- scale dependent\\
$ m_e, m_{\mu}, m_{\tau} $ &  $<$ 1/3\% &  high accuracy \\
$m_c, m_b$ & $\sim$ 4\% &  less certain \\
$m_u/m_d, m_s/m_d$ & ? & chiral Lagrangian -- ambiguous \\
$(m_u + m_d)/2$ & ? & QCD sum rules \\
$|V_{cd}| \approx |V_{us}|$ &  $\sim$ 1.5\% & fair accuracy\\
$|V_{cb}|, \left|V_{ub}/V_{cb}\right|$ & $\sim$ 20\% & poorly known \\
\hline
$m_t, m_H$ & $m_t = 150 \begin{array}{cccc}
+ & 19 & + & 15\\
- & 24 & - & 20 \end{array}$  &  $60 < m_H <
1000$  in SM \\ $m_t, m_H$ & $m_t = 131 \begin{array}{cccc}
+ & 23 & + & 5\\
- & 28 & - & 5 \end{array}$  &  $60 < m_H <
150$  in MSSM \\ $m_t$ &   $m_t > 108$ GeV  &  Fermilab data\\
\hline
$M_Z$ & $<$ .025\%  & high accuracy\\
 J (Jarlskog invariant) & $\sim 30\%$  &  uncertainty in $B_K$\\
\hline \end{tabular}
\end{center}
\end{table}

Some of these parameters are known to high accuracy, these include
$\alpha, \sin^2\theta_W, m_e,$ $m_{\mu}, m_{\tau}, M_Z$.  A few are
known with fair accuracy, $|V_{cd}|, m_c, m_b$.  Finally, the
following are poorly known:  $\alpha_s(M_Z), m_u, m_d, m_s, |V_{cb}|,
\left|V_{ub}/V_{cb}\right|, m_t, m_H$ and the Jarlskog invariant
measure of CP violation, J.

Our knowledge of $\alpha_s(M_Z)$ suffers predominantly from
theoretical uncertainties associated with renormalization scale
ambiguities.  The experimentally allowed range is now $\alpha_s(M_Z)
= .118 \pm .007$~$^3$,  where the errors probably underestimate the
theoretical uncertainties.

Light quark masses are determined using chiral Lagrangians and QCD
sum rules.  In the chiral Lagrangian approach, light fermion masses
are characterized by their transformation properties under the $SU(3)_L
\times SU(3)_R$  chiral symmetry. The mass matrix, $M$, transforms as
$(3,{\bar 3})$.  However, as shown by Kaplan and
Manohar~$^4$,   $M' = M + \kappa det(M) M^{\dagger -1}$ with
$\kappa$ an arbitrary constant, has the same transformation property.
This leads to a large theoretical uncertainty in light quark masses.

The weak mixing angles $|V_{ub}|$ and $|V_{cb}|$ suffer from both
large theoretical and/or experimental uncertainties.  The theoretical
uncertainties are evidenced by the different model dependent
calculations of B decay.  It is hoped that the application of heavy
quark effective field theory to the exclusive semileptonic B decays
will reduce these uncertainties, but this requires better
statistics.  The latest experimental result from $B \rightarrow D^* l
\nu$ is~$^5$  $|V_{cb}| = .050 \pm .008 \pm .007$ where the
first error is from statistics and the second is from extrapolation
uncertainties. The ratio $\left|V_{ub}/V_{cb}\right|$, determined
from inclusive B decay, has large model dependences.  Recently the
experimental results from CLEO II changed significantly from previous
measurements by both Argus and CLEO.  The latest results
give~$^5$ $.038 \le \left|V_{ub}/V_{cb}\right| \le .097$.  This
is a factor of two smaller than previous results and the change was
significantly larger than any of the experimental errors for any
particular model.  We will just have to wait and see if this settles
down with more data.

The top and Higgs masses are only known from the radiative effects
they have on electroweak parameters.  The latest data, analyzed
by Langacker~$^2$, gives the results presented in Table 1, where MSSM
denotes the minimal supersymmetric standard model. The difference
between the SM and MSSM result is the requirement of a lighter
Higgs boson in the MSSM.

Finally,  J suffers from strong interaction uncertainties associated
with the so-called bag constant, $B_K$.  The value of J$\times B_K$ can
be derived given the experimental value for $\epsilon_K$, all quark
masses, and the magnitudes of all Kobayashi-Maskawa [KM] elements.
Thus, the uncertainty in J is limited by the uncertainty in $B_K$ which
is of order $30\%$.

I have briefly reviewed the status quo, since some
of these parameters will be used as imputs to fix the fundamental
parameters in the fermion mass matrices and the others test the
theory.  For example, the biggest uncertainties in our prediction of
the top mass comes from the uncertainties in $m_b$ and
$\alpha_s(M_Z)$.  Reducing these uncertainties would make
this prediction much tighter.

\section{SUSY GUTs and the virtues of SO(10)}

$\bullet$  {\bf\it SUSY GUT}

Given two parameters the fine structure constant, $\alpha_G$, at the
GUT scale, $M_G$, one determines the three low energy
parameters, $\alpha, \sin^2\theta_W$ and $\alpha_s$, all evaluated at
some renormalization scale,$\mu$, which for convenience we might
choose to be $M_Z$.  In actuality, we use the two best
determined parameters, $\alpha$ and $\sin^2\theta_W$, to fix
$\alpha_G \sim 1/25$, $M_G \sim 10^{16} GeV$ and predict (in a SUSY
GUT) $\alpha_s(M_Z) = .125 \pm .002 \pm .009$ for a central value
of $m_t = 138 GeV.$~$^2$\footnote{In a non-SUSY GUT the prediction
is $\alpha_s(M_Z) \sim .07 $~$^2$, which is inconsistent with the
data.}   The errors take into account uncertainties in $m_H, m_t$ and
estimates of threshold corrections at both $M_G$ and the weak scale.
This result is in remarkably good agreement with the data. For a
heavier top, the central value for $\alpha_s$ decreases by several
percent.  This prediction assumes a supersymmetric desert, i.e.  the
only threshold between $M_Z$ and $M_G$ occurs below $\sim$1 TeV due to
the new spectrum of states encountered in the MSSM.

In any GUT, the number of fundamental parameters in the gauge
sector of the theory decreases by one and only in a SUSY GUT does
the resulting prediction agree with the low energy data [LED].  The
powerful assumption of a SUSY desert has an equally important
consequence. The LED becomes a window into physics at the GUT scale,
i.e. measurements at the weak scale gives us information about the
physics at $M_G$.

$\bullet$  {\bf\it Virtues of SO(10)}

\begin{description}
\item[] A single family of fermions fits into one irreducible
representation --- i.e.  $16_i \supset \{ u_i, d_i, e_i, \nu_i\} ~{\rm
with} ~~i = 1,2,3 $ labelling the three families.  We take the 3rd
family to be the top family.
\item[] The two Higgs doublets required in the MSSM fit into one
irreducible representation  --- i.e. $10 \supset \{ H, H', H_3, H'_3\}$
where $H,H'$ are weak doublets necessary for weak symmetry breaking
and giving masses to quarks and leptons and $H_3, H'_3$ are color
triplet Higgs which must get mass of order $M_G$ to avoid rapid
proton decay.  We will return to this point shortly.
\end{description}

In order to accomplish the GUT scale symmetry breaking we must have
additional representations including $\{ 45, 16, \overline{16}, \cdots
\}$, where, for example, a $16$ and $\overline{16}$ vev can break SO(10)
to SU(5) and the $45$ vev can then break SU(5) to SU(3) $\times$ SU(2)
$\times$ U(1).  This may happen at the same scale, $M_G << M_P$, or at
two separate scales, where the first occurs at a scale $v_{10}$ such
that  $v_5 = M_G << v_{10} << M_P$.

We have not considered other possible representations
which may be relevant for GUT symmetry breaking, such as $54,
{}~126$, etc. We shall  now assume that only the $45$ plays a crucial role
in the generation of fermion masses.  It is thus necessary to elaborate
the possible directions the $45$ vev may point in the two dimensional
space of U(1) subgroups of SO(10) which  commute with SU(3) $\times$
SU(2) $\times$ U(1).  Although there are only two orthogonal directions
in this space, we nevertheless consider the following 4 possible
symmetry breaking vevs ---
\begin{eqnarray}
\langle 45 \rangle_X &=& v_{10} e^{i \alpha_X} {\bf X} \nonumber \\
\langle 45 \rangle_Y& =& v_{5} e^{i \alpha_Y} {\bf Y}
\end{eqnarray}
\begin{eqnarray}
\langle 45 \rangle_{B-L} &= &v_{5} e^{i \alpha_{(B-L)}} {\bf (B-L)}
\nonumber \\
\langle 45 \rangle_{T_{3R}} &= &v_{5} e^{i \alpha_{T_{3R}}}
{\bf {T_{3R}}}
 \end{eqnarray}
where we have explicitly represented them as two groups of two
orthogonal vevs.  We consider all four since two of these vevs (one
in each group) are well motivated.

${\bf X}$, in Eq. (1), is the U(1) direction which leaves SU(5)
invariant.  This is why we have taken the magnitude of the vev to be
$v_{10}$, whereas all the others are taken to be $v_5$ since they do
not commute with SU(5).

${\bf B-L}$, in Eq. (2), just measures baryon number minus lepton
number.  It can play a crucial role in splitting the weak doublet and
color triplet Higgs multiplets, i.e. solving the hierarchy problem.
The Higgs doublets carry zero B-L whereas the triplets have non-zero
B-L.  Thus if the Higgs in the $10$ gets mass by coupling to this $45$,
only the color triplets will acquire mass  at the scale $M_G$.  Hence,
this vev is expected to be a necessary ingredient in any complete
SO(10) model which also solves the hierarchy problem.

\section{Dynamic Principles}

Let us now discuss the dynamical principles which guide us towards a
theory of fermion masses.
\begin{description}
\item[0.] At zeroth order, we work in the context of a SUSY GUT with
the MSSM  below  $M_G$.
\item[1.] We use SO(10) as the GUT symmetry with three families of
fermions $\{ 16_i  ~~i = 1,2,3 \}$ and the minimal electroweak Higgs
content in one $10$.  Using SO(10) symmetry relations allows us to
reduce the number of fundamental parameters.
\item[2.] We will assume that there are family symmetries which enforce
zeros of the mass matrix,  although we will not specify these
symmetries at this time.  As we will make clear shortly, these
symmetries will be realized at the level of the fundamental theory
defined at $M_P$.
\item[3.] Only the third generation obtains mass via
a dimension 4 operator.  The fermionic sector of the Lagrangian thus
contains the term ${\cal L}_f \supset A ~O_{33} \equiv A ~~16_3 ~10
{}~16_3$.  This term gives mass to  t, b and $\tau$.  It results in the
symmetry relation --- $\lambda_t = \lambda_b = \lambda_{\tau} \equiv A
$ at $M_G$.   This relation has been studied before by Ananthanarayan,
Lazarides and Shafi [ALS]~$^6$ and using $m_b$ and $m_{\tau}$ as
input it leads to reasonable results for $m_t$ and $\tan \beta$.
\item[4.] All other masses come from operators with dimension $> 4$.
As a consequence,  the family hierarchy will be related to the ratio
of scales above $M_G$.  We will show shortly how to understand the
higher dimension operators in terms of an effective field theory at
$M_G$, obtained by integrating out states with mass $> M_G$.
\item[5.]  [{\bf Predictivity requirement}] ~We demand the
\underline{minimal set} of effective fermion mass operators at $M_G$
\underline{consistent with the {\bf LED}}.
\end{description}

Let us now consider the general {\bf operator basis for fermion
masses}.  Let ${\cal L}_{fermion}$ include operators of the form
\begin{equation}
{\bf O_{ij}} = {\bf 16_i} ~( \cdots )_n ~{\bf 10} ~( \cdots )_m ~{\bf
16_j}
\end{equation}
where
$$( \cdots )_n =  {M_G^k ~45_{k+1} \cdots 45_n \over M_P^l
{}~45_X^{n-l}}
$$
and the $45$ vevs in the numerator can be in any of the 4 directions,
${\bf X, Y, B-L, T_{3R}}$ discussed earlier.

We said that such operators can be considered as the result of
integrating out states with mass $> M_G$.  For example, you can
convince yourself that an operator of the form  $O_{22} =  16_2 ~10
{}~{45_{B-L} ~M_G \over 45_X^2} ~16_2$ is generated by the tree graph of
Figure 1, assuming $v_{10} >> v_5 \sim M_G$.  Note it is at this level
in the fundamental theory at $M_P$ that additional family symmetries
are needed to enforce zeros in the mass matrix.
 \begin{figure}
\vspace{1in} \caption{Tree diagram which leads to effective operator
when massive states are integrated out.}
\end{figure}
It is also trivial to evaluate the
Clebsch-Gordon coefficients associated with any particular operator
since the matrices $X,Y,B-L,T_{3R}$ are diagonal.  Their eigenvalues on
the fermion states are given in Table 2.

\begin{table}[t]
\begin{center}
\begin{tabular}{|c|c|c|c|c|}
\multicolumn{5}{l}{Table~2. Quantum numbers of the}\\
 \multicolumn{5}{l}{four 45 vevs on fermion states.}  \\
\multicolumn{5}{l}{Note, if $u$ denotes a left-handed}\\
\multicolumn{5}{l}{up quark, then ${\bar u}$ denotes } \\
\multicolumn{5}{l}{a left-handed charge conjugate }\\
\multicolumn{5}{l}{up quark. }\\
\multicolumn{1}{c}{}&\multicolumn{4}{c}{}\\ \hline\hline
& ${\bf X}$ & ${\bf Y}$ & ${\bf B-L}$ & ${\bf T_{3R}}$
\\ \hline $u$ &  1 &  1/3 & 1 & 0  \\
${\bar u}$ &  1 & -4/3 & -1 & -1/2 \\
$d$ & 1 & 1/3 & 1 & 0 \\
${\bar d}$ & -3 & 2/3 & -1 & 1/2 \\
$e$ & -3 & -1 & -3 & 0 \\
${\bar e}$ & 1 & 2 & 3 & 1/2 \\
$\nu$ & -3 & -1 & -3 & 0 \\
${\bar \nu}$ & 5 & 0 & 3 & -1/2\\ \hline
\end{tabular}
\end{center}
\end{table}

\section{Operator analysis}

Our goal is to find the {\em minimal} set of fermion mass
operators consistent with the LED.  With any given operator set
one can evaluate the fermion mass matrices for up and down quarks and
charged leptons.  One obtains relations between mixing angles and
ratios of fermion masses which can be compared with the data.  It is
easy to show, however, without any detailed calculations that the
minimal operator set consistent with the LED is given by
\begin{eqnarray}
{\cal L}_{fermion} \supset & O_{33} + O_{23} + O_{22} + O_{12}&  ---
``22" ~{\rm texture} \nonumber\\
 {\rm or} & &  \\
&  O_{33} + O_{23} + O'_{23} + O_{12}& --- ``23'" ~{\rm texture}
\nonumber
\end{eqnarray}

It is clear that at least 3 operators are needed to give
non-vanishing mass to all charged fermions, i.e. $ ~det (m_a)  \neq 0$
for $a = u,d,e$.  That the operators must be in the [33, 23 and 12]
slots is not as obvious but is not difficult to show.  It is then
easy to show that 4 operators are required in order to have  CP
violation.  This is because, with only 3 SO(10) invariant operators,
we can redefine the phases of the three 16s of fermions to remove the
three arbitrary phases.  With one more operator, there is one
additional phase which cannot be removed.  A corollary of this
observation is that this minimal operator set results in just 5
arbitrary parameters in the Yukawa matrices of all fermions,  4
magnitudes and one phase.  This is the minimal parameter set which
can be obtained without solving the remaining problems of the fermion
mass hierarchy, one overall real mixing angle and a CP violating
phase.  We should point out however that the problem of understanding
the fermion mass hierarchy and mixing has been rephrased as the problem
of understanding the hierarchy of scales above $M_G$.  Moreover given
any particular operator set which fits the low energy data we would
be obliged at some later time to construct the complete GUT theory,
including symmetries which forbid additional operators and a
consistent description of symmetry breaking scales.  We leave this
problem for future analysis.

For now we shall describe the detailed analysis of the ``22"
texture.\footnote{As of this writing, we have found no models with
``23'" texture which fit the LED.}  Models with ``22" texture
give the following Yukawa matrices at $M_G$ --

\begin{equation}
{\bf \lambda_a} = \left( \begin{array}{ccc}
0 & z_a ~C & 0\\
z'_a ~C & y_a ~E ~e^{i \phi} & x_a ~B\\
0 & x'_a ~B & A
\end{array} \right)
\end{equation}
 with the subscript $a = \{ u, d, e\}$.  The constants
$x_a, x'_a, y_a, z_a, z'_a$ are Clebschs which can be determined once
the 3 operators ( $O_{23},O_{22}, O_{12}$) are specified.  Recall, we
have taken $O_{33} = A ~16_3 ~10 ~16_3$, which is why the Clebsch in
the 33 term is independent of $a$. Finally, combining the Yukawa
matrices with the Higgs vevs to find the fermion mass matrices we have
6 arbitrary parameters given by $A, B, C, E, \phi$ and $\tan \beta$
describing 14 observables.  We thus obtain 8 predictions.  We shall
use the best known parameters, $e, \mu, \tau, c, b, |V_{cd}|$, as
input to fix the 6 unknowns.  We then predict the values of $u, d, s,
t, \tan \beta, |V_{cb}|, |V_{ub}|$ and $J$.

We now show how (within the context of our dynamic principles) we can
use the LED to guide us towards {\em the theory of fermion masses} at
$M_G$.  We search for all operators ( $O_{23}, O_{22}, O_{12}$) with
dimension $\le 10$ which fit the data.  Using a coarse-grained
analysis,  it is easy to show that just 9 models (with one caveat) may
fit the data.  We now describe this analysis.

\subsection{\it 3rd generation fit --- $O_{33}$}

Using the values of $\alpha$ and $\sin^2\theta_W$ evaluated at $M_Z$,
we obtain $\alpha_G, M_G$ and $\alpha_s(M_Z)$.  There is a
theoretical uncertainty in this prediction due to unknown threshold
corrections at both $M_G$ and the weak scale.  There is also a 10\%
experimental uncertainty in $\alpha_s(M_Z)$.  Using arbitrary
threshold corrections we can obtain any experimentally allowed
value of $\alpha_s$.  We thus allow for all values of $\alpha_s(M_Z) =
.12 \pm .01$ self-consistently by introducing arbitrary threshold
corrections.

The analysis for the third generation follows ---
\begin{eqnarray}
m_b = {v \over \sqrt{2}} ~A ~\cos \beta ~\left( {\eta_b \over
S_b}\right) \\
m_{\tau} = {v \over \sqrt{2}} ~A ~\cos \beta ~\left( {\eta_{\tau} \over
S_{\tau}}\right) \\
m_t = {v \over \sqrt{2}} ~A ~\sin \beta \left( {1 \over
S_t}\right)
\end{eqnarray}
where  $v = 246 GeV$ and the terms in parentheses are renormalization
group factors which are implicit functions of both $A$ and
$\alpha_s(M_Z)$.  The numerator takes into account renormalization
from $M_Z$ to $m_b ~{\rm or} ~m_{\tau}$, and the denominator takes into
account the running from $M_G$ to $M_Z$.  We use two loop
renormalization group equations.

Using $m_b = 4.25 \pm .1 GeV$ and $m_{\tau} = 1.7841 GeV$
(these are running masses $m(m)$) as input for a given value of
$\alpha_s = .118$(for example) we obtain $A$ from the relation
$$
{m_b \over m_{\tau}} =  \left( {\eta_b \over \eta_{\tau}} {S_{\tau}
\over S_b} \right)(\alpha_s(M_Z), A) .
$$
Plugging this value of $A$ into the expression for $m_{\tau}$ we then
obtain $\tan \beta$.  We find $\tan \beta \sim 59$ for $m_b = 4.34
GeV$.  Now using the values of  $A$ and $\tan \beta$ in the expression
for $m_t$ we find $m_t(pole) \sim 188 GeV$.  In general, we find
values of $\tan \beta = 54 \pm 5$ and $m_t = 185 \pm 15 GeV$.  Both
$m_t$ and $\tan \beta$ increase for increasing values of $\alpha_s$ or
decreasing values of $m_b$.  Thus $m_t = 170 GeV$ and $\tan \beta = 50$
is obtained for $\alpha_s(M_Z) = .110, ~m_b = 4.35 GeV$.  Similar
results have been obtained previously by ALS~$^6$.  Note they find
lower values of $m_t$ since they allow for values of $\alpha_s$ which
are lower than those presently admissable by the data.

\subsection{\it 2nd generation --- $O_{22}$}

Let us now consider the 2nd generation.  We have 4 relations which
must be satisfied by the LED.
\begin{eqnarray}
|V_{cb}| \approx  |x_u - x_d| ~{B \over A} \sim 1/20 \\
{m_{\mu} \over m_{\tau}} \approx |y_e ~{E \over A} ~e^{i \phi} - x_e
{}~x'_e {B^2 \over A^2}| \sim 1/17 \\
{m_s \over m_b} \approx |y_d ~{E \over A} ~e^{i \phi} - x_d ~x'_d
{B^2 \over A^2}| \sim 1/25 \\
{m_c \over m_t} \approx |y_u ~{E \over A} ~e^{i \phi} - x_u ~x'_u
{B^2 \over A^2}| \sim 10^{-2}
\end{eqnarray}

We have written these equations using the parameters at $M_G$,
ignoring for the moment small renormalization group corrections.
Using the first relation we see that the ratio $B/A \sim 1/10$,
assuming Clebschs of order 1.  The 2nd and 3rd relations thus require
$E/A \sim 1/10$.  The last relation  then requires that the
Clebsch, $y_u << 1$.  Finally, the relation $m_s = m_{\mu}/3$ at
$M_G$, first suggested by Georgi-Jarlskog~$^7$, must be
incorporated, since it is in good agreement with the LED.  We thus
conclude that the Clebschs, $y_a$,(including RG corrections) should
approximately be in the ratio\footnote{If the Clebschs are not of
order 1 as we assumed, then it is possible that the 4 relations may be
satisfied with some fine-tuning and a completely different ratio of
Clebschs.  We have not pursued this possibility further. This is our
one caveat.}  $$ y_u : y_d : y_e = <1/3 : 1 : 3 . $$

The Clebschs, $y_a$, are derived from the operator $O_{22}$.  We have
searched over all dimension 5 and 6 operators to find
solutions to the above Clebsch ratios.  We find 6 solutions ---
\begin{eqnarray}
16_2 ~(45_X) ~10 ~({45_{B-L} \over 45_X}) ~16_2  \\
16_2 ~({1 \over 45_X}) ~10 ~(45_{B-L}) ~16_2 \nonumber \\
16_2 ~(45_X) ~10 ~(45_{B-L}) ~16_2 \nonumber \\
16_2 ~10 ~({45_{B-L} \over 45_X}) ~16_2 \nonumber \\
16_2 ~10 ~(45_{X} ~45_{B-L}) ~16_2 \nonumber \\
16_2 ~10 ~({45_{B-L} \over 45_X^2}) ~16_2 \nonumber
\end{eqnarray}

However, note that {\em all} solutions give the {\em same} ratio of
Clebschs --- \begin{equation}
 y_u : y_d : y_e = 0 : 1 : 3 .
\end{equation}

\subsection{\it 1st generation  --- $O_{12}$}

We can now show that the operator  $O_{12}$ is {\em unique}.  The first
two generations satisfy the relations ---
\begin{eqnarray}
{m_d \over m_s} \approx 9 ~{z_d ~z'_d \over z_e ~z'_e}~{m_e \over
m_{\mu}} \left( {\eta_{\mu} ~\eta_d \over \eta_e ~\eta_s} \right) \\
{m_u \over m_d} \approx {z_u ~z'_u \over z_d ~z'_d}~{m_s \over
m_c} ~\tan^2 \beta ~\left( {S_d \over S_u} \right)^2 \\
|V_{cd}| = \left| \sqrt{{z_d \over z'_d}}\sqrt{{m_d \over m_s}} -
\sqrt{{z_u \over z'_u}}\sqrt{{m_u \over m_c}} ~e^{-i \phi} \right|
\end{eqnarray}

The first relation is satisfied {\em if} $$z_d ~z'_d \approx z_e ~z'_e
.$$  This relation is satisfied if the Clebsch is derived from an SU(5)
invariant vev, i.e.  $45_X$.  Since $\tan \beta$ is large, the second
relation requires
$${z_u ~z'_u \over z_d ~z'_d} \approx ({1 \over 3})^{6~{\rm or}~7}.$$
Finally, the last relation requires $$z_d \approx z'_d .$$

The unique operator which satisfies the above 3 relations is
\begin{equation}
 O_{12} = 16_1 ~({45_X \over M_P})^3 ~10 ~({45_X \over
M_P})^3 ~16_2 .
\end{equation}

\subsection{\it $O_{23}$}

We have now determined, using simple arguments, all but one of the
operators.  The charged fermion mass matrices are given by ---
\begin{eqnarray}
U& = & \left( \begin{array}{ccc} 0 & C & 0 \\
                                C & 0 & x_u ~B \\
                                0 & x'_u ~B & A \end{array} \right)\\
D& = &\left( \begin{array}{ccc} 0 & - 27 ~C & 0 \\
                                - 27 ~C & E ~e^{i \phi} & x_d ~B \\
                                0 & x'_d ~B & A \end{array} \right) \\
E& =  &\left( \begin{array}{ccc} 0 & - 27 ~C & 0 \\
                                - 27 ~C & 3 ~E ~e^{i \phi} & x_e ~B \\
                                0 & x'_e ~B & A \end{array} \right)
\end{eqnarray}

With this form for the mass matrices, the KM element $V_{cb}$ satisfies
the relation ---
\begin{equation} |V_{cb}| \approx  \chi ~\sqrt{{m_c \over m_t}}
{}~\sqrt{{S_u \over \eta_c ~S_t ~S}} \sim  .058  ~\chi .
\end{equation}
where $\chi \equiv  { |x_u - x_d| \over \sqrt{|x_u x'_u|}}$ and the
last term results from using central values of the input parameters.

Experimentally, we have an upper bound on $|V_{cb}| \le .054$.  We thus
require that the function of Clebschs, $\chi$, satisfy  $\chi < 1$.

We have searched over all dimension 5 and 6 operators for $\chi < 1$.
We find 9 possible models with only 3 different values of $\chi = {2
\over 3}, ~{5 \over 6},{\rm or} ~{8 \over 9}$.  The 9 models are listed
below.

$$\chi = 2/3$$
\begin{eqnarray}
1 & & 16_2 ~(45_Y) ~10 ~({1 \over 45_X}) ~16_3  \\
2 & & 16_2 ~(45_Y) ~10 ~({45_{B-L} \over 45_X}) ~16_3  \\
3 & & 16_2 ~({45_Y \over 45_X}) ~10 ~({1 \over 45_X}) ~16_3 \\
4 & & 16_2 ~({45_Y \over 45_X}) ~10 ~({45_{B-L} \over 45_X}) ~16_3
\end{eqnarray}

$$\chi = 5/6$$
\begin{eqnarray}
5 & & 16_2 ~(45_Y) ~10 ~({45_Y \over 45_X}) ~16_3  \\
6 & & 16_2 ~({45_Y \over 45_X}) ~10 ~({45_Y \over 45_X}) ~16_3
\end{eqnarray}

$$\chi = 8/9$$
\begin{eqnarray}
7 & & 16_2 ~10 ~({1 \over 45_X^2}) ~16_3  \\
8 & & 16_2 ~10 ~({45_{B-L} \over 45_X^2}) ~16_3  \\
9 & & 16_2 ~10 ~({45_{B-L} \over 45_X})^2 ~16_3
\end{eqnarray}

This is as far as we can get with our coarse-grained search.  We
must now take the 9 distinct models and test the predictions.
We are presently in the midst of a complete renormalization group
analysis, obtaining predictions as a function of the input
parameters, $e, \mu, \tau, c, b, |V_{cd}|$ and $\alpha_s(M_Z)$.
For now we present some preliminary results.

\section{Preliminary results}

The results in Table 3 are for inputs $m_c = 1.23 GeV, m_b = 4.34
GeV, |V_{cd}| = .221$ and $\alpha_s(M_Z) = .118$ for 4 different
models.  The model numbers are defined in Eqs. (23 - 31)

\begin{table}[h]
\begin{center}
\begin{tabular}{|c|c|c|c|c|}
\multicolumn{5}{l}{Table~3. Results for models 3, 6, 8, and
9.}  \\
\multicolumn{5}{c}{}\\ \hline\hline & 3 & 6 & 8 & 9  \\ \hline
$m_d [MeV]$ & 6.8 &  6.9 & 6.9 & 7.3  \\
 $ m_u/m_d$ &  .82 & .79 & .80 & .79 \\
$m_s/m_d$ & 24.8 & 24.1 & 24.9 & 22.4 \\
$m_t$ & 188 &188 & 188 &188 \\
$\tan \beta$ &59 & 59 & 59 & 59 \\
$|V_{cb}|$ & .039 & .049 & .052 & .052 \\
$\left| V_{ub}/V_{cb}\right|$ & .065 &.064 & .066 & .068 \\
$J \times 10^5$ & 1.8 & 2.9 & 3.3 & 3.7 \\
 \hline
\end{tabular}
\end{center}
\end{table}

Note,  $|V_{cb}| ~{\rm and} ~J$ are sensitive to the value of $\chi$
and so can distinguish between the 3 types of models.  With better
data, $|V_{cb}|$ will distinguish between these choices.  At present we
can use $\epsilon_K$, the CP violating parameter in $K$ decay, to
distinguish these models through its dependence on $J$.  We
find that models with $\chi = 2/3$ tend to give too little CP
violation.  Finally, we see that $m_s/m_d ~{\rm and} ~m_u/m_d$ will
also be useful in constraining the models.  Complete results will be
presented in an upcoming publication.~$^1$

\section{Conclusion}

We have presented a straightforward method for trying to understand the
origin of fermion masses.  There are four main ingredients ---
\begin{itemize}
\item The assumption of a SUSY GUT with its SUSY DESERT implies that
the Low Energy Data becomes a window into the physics at the GUT
scale.

\item The assumption of symmetries (SO(10) plus family symmetries)
with a general operator analysis allows us to reduce the number of
fundamental parameters in the theory.

\item With the above assumptions, we obtain predictive theories of
fermion masses.  Given 6 inputs, $e, \mu, \tau, c, b,$ $|V_{cd}|$, we
find 8 predictions for $u,d,s,t,\tan\beta, |V_{cb}|, $ $|V_{ub}|$ and
$J$.

\item This method allows us to use the LED to systematically find the
theory of fermion masses at $M_G$.
\end{itemize}

A final note:  there is no apriori reason to believe that this scheme
should work, {\em but} if it does work, i.e. if our results agree
with the LED,  then perhaps we will have learned something about the
physics at the GUT scale.  For example, consider the set of possible
models 1 -- 9 in Eqs. (23  - 31).  Each one has the vev $45_X$ in
the denominator.  This can only occur if this vev gives mass $ >> M_G$
to some fermion. When this fermion is integrated out in the effective
theory at $M_G$, the vev then appears in the denominator of these
higher dimension operators.  Thus fermion masses and mixing angles at
low energies may tell us about the hierarchy of symmetry breaking
scales above $M_G$.

\end{document}